\documentclass[a4paper,11pt]{article}
\usepackage[utf8]{inputenc}
\usepackage{amsfonts}
\usepackage{amssymb}
\usepackage{graphics}
\usepackage{fancyhdr}
\usepackage{fancyvrb}
\usepackage{fancybox}
\usepackage{float}
\usepackage{geometry}
\usepackage{minitoc}
 \usepackage{indentfirst}
 \usepackage{multicol}
 \usepackage[outerbars]{changebar}
 \usepackage{pifont,textcomp}
 \usepackage{amsfonts}
 \usepackage{amsmath}
 \usepackage{amscd}
 \usepackage{array}
\usepackage{multirow}
 \usepackage{mathrsfs}
 \usepackage{amssymb}
 \usepackage{amsthm}

\title{Hilbert space representation for   maximal length and  minimal momentum uncertainties }
\title{Hilbert space representation of  maximal length and  minimal momentum uncertainties }
\author{ Kossi Amouzouvi$^{1}$,\, Benjamin A. Appiah$^{2}$\, Lat\'evi M. Lawson$^{3,4}$\, and\\  Abdel-Baset A. Mohamed$^{5,6}$\\ 
	\space\\
	${}^{1}\!$ Department of Mathematics,\\ Kwame Nkrumah University of Science and Technology,\\    Kumasi, Ghana \\\\
	${}^{2,3}\!$ African Institute for Mathematical Sciences (AIMS) Ghana\\
	Summerhill Estates, East Legon Hills, Santoe, Accra\\
	P.O. Box LG DTD 20046, Legon, Accra, Ghana\\\\
	${}^{4}\!$ Universit\'{e} de Lom\'{e}, Facult\'{e} des Sciences, D\'{e}partement de Physique,\\
	 Laboratoire de Physique des Mat\'{e}riaux et des Composants\\
	 à Semi-Conducteurs, 01 BP 1515 Lom\'{e}, Togo\\\\
	${}^{5}\!$ Department of Mathematics, College of Science and Humanities,\\ Prince Sattam Bin  Abdulaziz University, Al-Aflaj, Saudi Arabia \\\\
	 ${}^{6}\!$ Department of Mathematics, Faculty of Science, \\Assiut University, Assiut, Egypt \\\\
	Kossi@aims.edu.gh$^{1}$, bappiah@aims.edu.gh$^{2}$,  latevi@aims.edu.gh$^{3,4}$ and\\ abdelbastm@aun.edu.eg$^{5,6}$}

\begin{document}
\maketitle

\begin{abstract}
	
   	Perivolaropoulos has recently  proposed a position-deformed Heisenberg algebra which includes a maximal length  [Phys. Rev. D  95, 103523 (2017)]. He has shown that  this  length scale naturally emerges  in the context of cosmological particle’s horizon  or cosmic topology. Following this work,  we propose a new deformed algebra and  derive the  maximal length  uncertainty  and its corresponding minimal momentum uncertainty from the  generalized uncertainty principle. We construct a Hilbert  space representation in the  spectral representation of this length scale. We also construct the corresponding Fourier transform and its inverse representations. Finally, we propose $n$-dimensional representation  of this algebra.

\end{abstract}


\section{Introduction}

A minimum length scale of the order of Planck length is a feature of many models of quantum gravity that seek to unify quantum mechanics and gravitation. Recently, Perivolaropoulos \cite{1} proposed a new version of a Generalized Uncertainty
Principle (GUP)  that yields  simultaneously   a maximal length and a minimal momentum uncertainties. An extension of this work has been  carried out \cite{3} and some applications  
have been performed to investigate the implications of such
GUPs in thermodynamics of ideal gases \cite{4,5} and of Black hole \cite{6}.  In this work \cite{1}, Perivolaropoulos also predicted the simultaneous existence of minimal and
maximal length measurements  More recently, one of us has shown that both measurable lengths can be obtained from  position-dependent noncommutativity \cite{6'}. Its applications run from  quantum optics \cite{7}, quantum thermodynamics \cite{8} to  non-Hermitian quantum mechanic scenarios \cite{9}.

In the present letter, we consider a generalized version of  position-dependent deformed Heisenberg algebra \cite{1} which includes a simultaneous presence of maximal position and minimal momentum uncertainties. Based on the GUP with  minimal length scenarios \cite{10,11,12,13,14,15,16}, we study the  functional analytic aspects
of the maximal  length uncertainty. We show that the spectral representation of this length scale forms a family of discrete eigenvalues that describes  a lattice space. The  corresponding  set of eigenvectors  exhibits proprieties similar to the standard Gaussian states which are consequences of quantum fluctuation at this scale.
The rest of this letter   is structured as follows: in the next
section, we review the GUP with a maximal length and a minimal momentum.  In the third section, we study the position  representation of these   uncertainty measurements, the  corresponding Fourier transform and its inverse. Then finally, we generalise this  representation in
an arbitrary dimensional Hilbert  space. In the last section we present our conclusion. 

\section{Position-deformed  Heisenberg algebra}
Let  $\mathcal{H}_{\tau}=\mathcal{L}^2(\mathbb{R})$ be the
 Hilbert space of square integrable functions. $  \mathcal{D}(\hat X)$ and  $\mathcal{D}(\hat P)$ are respectively the domains of operators $\hat X$ and  $\hat P$ maximally dense in   $\mathcal{H}$.
These operators $\hat X $ and  $ \hat P$    satisfy \cite{6',7,8}
\begin{eqnarray}\label{al1}
	{[\hat X,\hat P ]}=i\hbar (\mathbb{I}-\tau \hat X +\tau^2 \hat X^2),
\end{eqnarray}
where  $\tau \in \mathbb{R}_+^*$   is the GUP  deformed parameter \cite{10,15,16,17}. By  setting these operators  
\begin{eqnarray}\label{op}
	\hat X=\hat x,\quad 
	\hat P=(\mathbb{I}-\tau \hat x +\tau^2 \hat x^2)\hat p,
\end{eqnarray}
in terms of the  Hermitian operators $\hat x$ and $\hat p$  satisfy the  ordinary Heisenberg algebra 
$[\hat x,\hat p]=i\hbar$,
one recovers the relation (\ref{al1}). Let   $\phi(x)\in \mathcal{D}(\hat X)$ and  $ \phi(p)\in \mathcal{D}(\hat P)$ be respectively  the position and momentum representations. The  action  of the operators (\ref{op}) on these square integrable functions reads as follows
\begin{eqnarray}
	\hat X \phi(x)&=&x \phi(x) \quad \mbox{and}\quad  \hat P \phi(x)=-i\hbar(1-\tau  x +\tau^2  x^2)\frac{d}{dx}   \phi(x), \quad x\in \mathbb{R}, \label{3}\\
	\hat X \phi(p)&=&i\hbar\frac{d}{dp}  \phi(p) \quad \mbox{and}\quad  \hat P \phi(p)=\left(1-i\hbar\tau  \frac{d}{dp}  -\tau^2 \hbar^2 \frac{d^2}{dp^2} \right)p\phi(p), \quad p\in \mathbb{R}\label{k}.
\end{eqnarray}
For  both representations, the corresponding completeness relations are given by 
\begin{eqnarray}
	\int_{-\infty}^{+\infty}\frac{dx}{1-\tau  x +\tau^2  x^2}|x\rangle \langle x|&=&\mathbb{I}, \label{identity} \\ \int_{-\infty}^{+\infty}dp|p\rangle \langle p|&=&\mathbb{I}.
\end{eqnarray}
Consequently, the scalar product between two states $|\Psi\rangle$ and $|\Phi\rangle$ and the orthogonality of
eigenstates become
\begin{eqnarray}
	\langle \Psi |\Phi\rangle&=& \int_{-\infty}^{+\infty}\frac{dx}{1-\tau  x +\tau^2  x^2} \Psi^*(x)\Phi(x),\quad 
	\langle x |x'\rangle= (1-\tau  x +\tau^2  x^2)\delta(x-x'),\\
	\langle \Psi |\Phi\rangle &=& \int_{-\infty}^{+\infty}dp  \Psi^*(p)\Phi(p),\quad \quad  \quad \quad \quad \quad \quad  \langle p |p'\rangle= \delta(p-p').
\end{eqnarray}
For an operator $\hat A=\{\hat X,\hat P\}$, its expectation value   for both representations  are given by
\begin{eqnarray}
\langle \hat A  \rangle_{\phi(x)}&=&	\langle \phi(x)|\hat A |\phi(x) \rangle=\int_{-\infty}^{+\infty}\frac{dx}{1-\tau  x +\tau^2  x^2} \phi^*(x)\hat A\phi(x),\quad \\
\langle \hat A  \rangle_{\phi(p)}&=&	\langle \phi(p)|\hat A |\phi(p) \rangle=\int_{-\infty}^{+\infty}dp \phi^*(p)\hat A\phi(p),
\end{eqnarray}
 and the corresponding dispersions are 
\begin{eqnarray}
(\Delta_{\phi(x)} A)^2&=&	\langle \hat A^2  \rangle_{\phi(x)}-\langle \hat A \rangle_{\phi(x)}^2=\int_{-\infty}^{+\infty}\frac{dx}{1-\tau  x +\tau^2  x^2}\phi^*(x)\left( \hat A-\langle \hat A  \rangle_{\phi(x)}\right)^2\phi(x),\quad \\
(\Delta_{\phi(p)} A)^2	&=&	\langle \hat A^2  \rangle_{\phi(p)}-\langle \hat A \rangle_{\phi(p)}^2=\int_{-\infty}^{+\infty}dp \phi^*(p)\left( \hat A-\langle \hat A  \rangle_{\phi(p)}\right)^2\phi(p).
\end{eqnarray}

 For any representation,  an interesting feature  can be observed from  the commutation relation  (\ref{al1})  through the  following  uncertainty relation :
\begin{eqnarray}\label{uncertitude}
	\Delta  X\Delta P\geq \frac{\hbar}{2}\left(1-\tau\langle \hat X\rangle+\tau^2\langle \hat X^2\rangle\right).\label{in2}
\end{eqnarray}
Using the relation 
$ \langle \hat X^2\rangle=(\Delta  X)^2+\langle \hat X\rangle^2$, the equation (\ref{uncertitude}) can be rewritten as a second order equation for $\Delta X$ whose solution is 
\begin{equation}
	\Delta X=\frac{\Delta P}{\hbar \tau^2}\pm \sqrt{\left(\frac{\Delta P}{\hbar \tau^2}\right)^2
		-\frac{\langle \hat X\rangle}{\tau}\left(\tau\langle \hat X\rangle-1\right)
		-\frac{1}{\tau^2}}.
\end{equation}
This equation leads to the absolute minimal uncertainty $\Delta P_{min}$ in $P$-direction  and the absolute maximal uncertainty  $\Delta X_{max}$ in $X$-direction  
when $\langle  \hat X\rangle=0$, such that
\begin{eqnarray}
	\quad \Delta X_{max}=\frac{1}{\tau}\quad  \mbox{and}\quad
	\Delta P_{min}=\hbar\tau.
\end{eqnarray}
It is well known that \cite{10}, the   existence of minimal uncertainty   raises
the question of the loss  of representation  i.e the space is inevitably bounded by minimal quantity beyond which any further localization of particles is not possible. In the  presente situation, the minimal  momentum $\Delta P_{min}$   leads to the loss of representation in $P$-direction i.e the loss of $\phi(p)$-representation and a maximal measurement $\Delta X_{max}$  conversely will be the physical  space of  wavefunction representations i.e  all functions $\phi (x)\in  \mathcal{D}(\hat X) \subset  \mathcal{H}= \mathcal{L}^2(-\infty,+\infty)$   vanish  at the  boundary $ \phi(-\infty)=0=\phi (+\infty)$. Furthermore, the existence of the  minimal momentum $\Delta P_{min}$ induces also the loss of Hermicity of the operator $\hat P$ 
\begin{eqnarray}
	\hat P_y^\dag= \hat P_y+i\hbar\tau(\mathbb{I}-2\tau \hat X) \implies \hat P^\dag\neq \hat P.
\end{eqnarray}
Using the  relation (\ref{identity}) and by performing a partial integration \cite{6'}, one can easily show that 
\begin{eqnarray}
	\langle \psi|\hat P\phi\rangle  = \langle \hat P^\dagger \psi| \phi\rangle,
\end{eqnarray}
where 
\begin{eqnarray}
	\mathcal{D} (\hat P)&=&\{\phi,\phi'\in \mathcal{L}^2(-\infty,\,\infty); \phi (-\infty)=0=\phi' (+\infty) \}, \\ 
	\mathcal{D} (\hat P^\dag  )&=&\{\psi,\psi'\in \mathcal{L}^2(-\infty,\,\infty) \}.
\end{eqnarray}

\section{Representation  of the maximal length and the minimal momentum }
In this  section, we study the Hilbert  space representation of  the deformed Heisenberg algebra (\ref{al1}). We define the position space representation that describes the maximal length uncertainty and its  corresponding minimal momentum uncertainty. The quasi-representation of the position wavefunction is constructed through the Fourrier transform and its inverse. Finally  we
extend  the deformed Heisenberg algebra (\ref{al1}) to $n$-dimensional case.
\subsection{ Position space representation }
The representation of the momentum operator $\hat P$  reads as  follows
\begin{eqnarray}
	-i\hbar(1-\tau  x +\tau^2  x^2)\frac{d}{dx}   \phi_\eta(x)&=& \eta  \phi_\eta (x), \quad\quad \eta\in\mathbb{R}.\label{diff}
\end{eqnarray} 
By solving the  differential equation (\ref{diff}),
we obtain the  position eigenvectors in the  form 
\begin{eqnarray}\label{fzeta}
	\phi_\eta (x)= A\exp\left(i\frac{2\eta}{\tau\hbar \sqrt{3}}\left[\arctan\left(\frac{2\tau x-1}{\sqrt{3}}\right)
	+\frac{\pi}{6}\right]\right),
\end{eqnarray}
where $A$ is an abritrary constant.
Then by normalization, $\langle \phi_\eta|\phi_\eta\rangle=1$, we have 
\begin{eqnarray}
	1&=&\int_{-\infty}^{+\infty}\frac{dx}{1-\tau  x+\tau^2 x^2} \phi_\eta^* (x) \phi_\eta (x)\cr
	&=& A^2 \int_{-\infty}^{+\infty}\frac{dx}{1-\tau  x+\tau^2 x^2}.
\end{eqnarray}
so, we find
\begin{eqnarray}\label{nzeta}
	A&=& \sqrt{\frac{\tau\sqrt{3}}{2\pi}}.
\end{eqnarray}
Substituting this equation (\ref{nzeta}) into the equation (\ref{fzeta}), we get 
\begin{eqnarray} \label{nor}
	\phi_\eta (x) &=& \sqrt{\frac{\tau\sqrt{3}}{2\pi}} \exp\left(i\frac{2\eta}{\tau\hbar \sqrt{3}}\left[\arctan\left(\frac{2\tau x-1}{\sqrt{3}}\right)
	+\frac{\pi}{6}\right]\right).
\end{eqnarray}
This  wave function   describes  the maximal length  and the minimal momentum uncertainties. It is  well known that \cite{31'} a compression of the wave function to a small volume leads
to a sufficient increase of its kinetic energy, which leads to a  divergence of the wave function. Since the  wave function (\ref{nor}) describes a particle in a small region, we calculate the expectation value of its kinetic energy 
\begin{eqnarray}
	E&=& \frac{1}{2m}\langle \phi_\eta|\hat P^2|\phi_\eta \rangle\\
	&=& \frac{1}{2m}\int_{-\infty}^{+\infty} \phi_\eta^* (x) \left[-i\hbar(1-\tau  x+\tau^2 x^2)\frac{d}{dx}\right]^2\phi_\eta (x).
\end{eqnarray}
By solving this integral, we find
\begin{eqnarray}
	E=  \frac{\eta^2 }{2m}  < \infty.
\end{eqnarray}
This result is known as  energetic stability or atomic stability \cite{31'}.
In comparison with Kempf et al \cite{10}, the expectation value of the energy in our case does not
diverge. According to this formalism \cite{10}, any states that have a well-defined  minimal uncertainty measurement which is inside of a forbidden gap cannot have a finite energy. So that they cannot be accepted as physical states. Conversly to this formalism, here the energy is well defined therefore the states $\phi_\eta(x)$ are physically relevant although
they are defined inside of the forbidden gap $0 < \Delta P < \Delta P_{min} $.
Thus, these states are sufficiently quantized  with $\eta= \frac{\tau \hbar \sqrt{3}}{2}n  \,(n\in \mathbb{N})$  and  they form a set of orthogonal basis. 

\begin{figure}[htbp]
	\resizebox{0.95\textwidth}{!}{
		\includegraphics{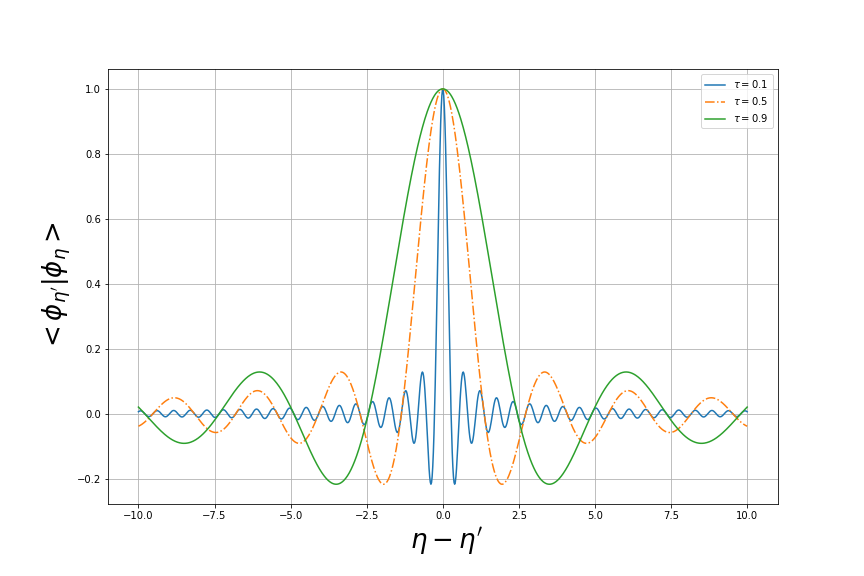}
	}
	\caption{\it \small Variation of $\langle \phi_{\eta'}|\phi_{\eta}\rangle$ versus $\eta-\eta'$.
	}
	\label{figwav}       
\end{figure}

The scalar product of the formal  eigenstates  is given by
\begin{eqnarray}
	\langle \phi_{\eta'}|\phi_{\eta}\rangle&=& \frac{\tau\sqrt{3}}{2\pi}\int_{-\infty}^{+\infty}\frac{dx}{1-\tau x+\tau^2 x^2}\cr&&\times\exp\left(i\frac{2(\eta-\eta')}{\tau\hbar \sqrt{3}}\left[\arctan\left(\frac{2\tau x-1}{\sqrt{3}}\right)+\frac{\pi}{6}\right]\right)\\
	&=& \frac{\tau\hbar\sqrt{3}}{2\pi(\eta-\eta')}\sin\left(2\pi\frac{\eta-\eta'}{\tau\hbar \sqrt{3}}\right).
\end{eqnarray}
This relation  shows that, the normalized  eigenstates (\ref{nor}) are  no longer
orthogonal. However,  if one tends $(\eta-\eta')\rightarrow \infty$, these states  become orthogonal 
\begin{eqnarray}
	\lim_{(\eta-\eta')\rightarrow \infty} \langle \phi_{\eta'}|\phi_{\eta}\rangle=0.\label{or}
\end{eqnarray}
For  $(\eta-\eta')\rightarrow 0$, we have 
\begin{eqnarray}
	\lim_{(\eta-\eta')\rightarrow 0} \langle \phi_{\eta'}|\phi_{\eta}\rangle=1.
\end{eqnarray}
These properties  show that, the  states  $|\phi_\eta\rangle$ are essentially Gaussians centered at $(\eta-\eta')\rightarrow 0$ (see Figure \ref{figwav}).  This indicates   quantum gravitational fluctuations  at this scale and these fluctuations increase as one increases the gravitational effects.

Since at  infinity  $(\eta-\eta'\rightarrow \infty)$, the scalar produt is zero, the states become orthogonal (\ref{or}). Moreover,  one has 
\begin{eqnarray}
	2\pi\frac{\eta-\eta'}{\tau\hbar \sqrt{3}}=n\pi\implies \eta-\eta'=\frac{\tau \hbar \sqrt{3}}{2}n, \quad n\in \mathbb{Z},
\end{eqnarray}
which  implies
\begin{eqnarray}
	\big\langle \phi_{(\eta+\frac{ \sqrt{3}}{2}n )\tau\hbar}\big|\phi_{(\eta+\frac{ \sqrt{3}}{2}n' )\tau\hbar}\big\rangle=\delta_{nn'}.
\end{eqnarray}
The set of eigenvectors $\big\{ \big|\phi_{(\eta+\frac{ \sqrt{3}}{2}n' )\tau\hbar}\big\rangle \big\}$ form a complete
orthogonal basis which can  diagonalize the operator $\hat P$. Since the formal momentum eigenvectors  $\big|\phi_{\frac{\tau \hbar \sqrt{3}}{2}n}\big\rangle $  are physically accepted and by referring  to the work  \cite{34}, one  may  tempted to interpret this result as if we are  describing physics on a
lattice in which   each  sites are spacing by the value $\frac{\sqrt{3}}{2}\Delta P_{min}. $

\subsection{  Fourier transform and its inverse representations }

Since the states $\phi_\eta(x)$ are  physically meaningful and are well localized,  one can obtain  the quasi-momentum representation  by projecting  an arbitrary state onto this  localized states \cite{10}
\begin{eqnarray}
	\psi (\eta)&= &\langle \phi_\eta|\psi  \rangle \cr
	&=&\sqrt{\frac{\tau\sqrt{3}}{2\pi}} \int_{-\infty}^{+\infty}\frac{\psi(x)dx}{1-\tau  x+\tau^2 x^2} 
	e^{-i\frac{2\eta}{\tau \hbar \sqrt{3}}\left[\arctan\left(\frac{2\tau x-1}{\sqrt{3}}\right)
		+\frac{\pi}{6}\right]}.\label{moment}
\end{eqnarray}
The transformation that maps the  quasi-position space wave functions into the momentum ones is the  Fourier transformation.
The inverse  Fourier transform is given by 
\begin{eqnarray}
	\psi(x)=\frac{1}{\hbar\sqrt{2\pi\tau\sqrt{3}}} \int_{-\infty}^{+\infty}d\eta \psi(\eta)  e^{i\frac{2\eta}{\tau \hbar \sqrt{3}}\left[\arctan\left(\frac{2\tau x-1}{\sqrt{3}}\right)
		+\frac{\pi}{6}\right]}.\label{F1}
\end{eqnarray}

Moreover  from equation (\ref{moment}), we can deduce
\begin{eqnarray}
	\frac{d}{d\eta} e^{-i\frac{2\eta}{\tau \hbar \sqrt{3}}\left[\arctan\left(\frac{2\tau x-1}{\sqrt{3}}\right)
		+\frac{\pi}{6}\right]}= -i\frac{2}{\tau \hbar \sqrt{3}}\left[\arctan\left(\frac{2\tau x-1}{\sqrt{3}}\right)
	+\frac{\pi}{6}\right]\cr\times e^{-i\frac{2\eta}{\tau \hbar \sqrt{3}}\left[\arctan\left(\frac{2\tau x-1}{\sqrt{3}}\right)
		+\frac{\pi}{6}\right]}.
\end{eqnarray}
This equation is equivalent to 
\begin{eqnarray}
	i\frac{\tau\hbar \sqrt{3}}{2}\frac{d}{d\eta}&=& \left[\arctan\left(\frac{2\tau x-1}{\sqrt{3}}\right)
	+\frac{\pi}{6}\right]\cr
	&=& \left[\arctan\left(\frac{2\tau x-1}{\sqrt{3}}\right)
	+\arctan\left(\frac{1}{\sqrt{3}}\right)\right].\label{eq}
\end{eqnarray}
From the following relation \cite{34}
\begin{eqnarray}
	\arctan\alpha +\arctan \beta=\arctan \left(\frac{\alpha+\beta}{1-\alpha\beta}\right) \quad \mbox{with} \quad \alpha\beta<1,
\end{eqnarray}
we deduce that
\begin{eqnarray}
	\tan \left[\arctan\left(\frac{2\tau x-1}{\sqrt{3}}\right)
	+\arctan\left(\frac{1}{\sqrt{3}}\right)\right]=\frac{\tau x \sqrt{3} }{2-\tau x}.
\end{eqnarray}
We notice from  the equation (\ref{eq}) that the position operator   is represented  as \cite{11}
\begin{eqnarray}
	\hat X&=&\frac{2}{\tau}\frac{\tan\left(i\frac{\tau\hbar \sqrt{3}}{2}\frac{d}{d\eta}\right)}{\sqrt{3}+\tan\left(i\frac{\tau\hbar \sqrt{3}}{2}\frac{d}{d\eta}\right)},\\
	\hat X\psi(\eta)&=&\frac{2}{\tau}\frac{\tan\left(i\frac{\tau\hbar \sqrt{3}}{2}\frac{d}{d\eta}\right)}{\sqrt{3}+\tan\left(i\frac{\tau\hbar \sqrt{3}}{2}\frac{d}{d\eta}\right)}\psi(\eta).
\end{eqnarray}
From the action of $\hat P$ on the  quasi representation (\ref{F1}) and using  equation (\ref{3}), we have
\begin{eqnarray}
	\hat P\psi(\eta)=\eta  \psi(\eta).
\end{eqnarray}
Note that in the limit of $\tau$, we
recover the corresponding ordinary quantum mechanics
results in momntum space   (\ref{k}) 
\begin{eqnarray}
\lim_{\tau\rightarrow 0}\hat X\psi(\eta)=i\hbar\frac{d}{d\eta}\psi(\eta),\quad \quad
\lim_{\tau\rightarrow 0}\hat P\psi(\eta)= \eta\psi(\eta).
\end{eqnarray}

Now, based on the representation (\ref{moment}) the  scalar product of two arbitrary vectors $|\Psi\rangle$ and $|\Phi\rangle$ reads  as follows
\begin{eqnarray}
	\langle \Psi|\Phi\rangle&=&\frac{1}{2\pi\hbar^2\sqrt{3}}\int_{-\infty}^{+\infty}\int_{-\infty}^{+\infty}\int_{-\infty}^{+\infty}\frac{1}{1-\tau  x+\tau^2 x^2}\cr&&
	e^{i\frac{2(\eta-\eta')}{\tau \hbar \sqrt{3}}\left[\arctan\left(\frac{2\tau x-1}{\sqrt{3}}\right)
		+\frac{\pi}{6}\right]}\Psi^*(\eta')\Phi(\eta) dxd\eta'd\eta.
\end{eqnarray} 

\subsection{Generalization to n-dimensional position-deformed algebra}
A generalization of the one-dimensional  deformed-position  Heisenberg algebra (\ref{al1}) that preserves the
rotational symmetry is
\begin{eqnarray}\label{algn}
	{[ {\bf \hat X}_i,  {\bf \hat P}_j ]}=i\hbar \delta_{ij} (\mathbb{I}-\tau {\bf \hat X } +\tau^2 {\bf \hat X^2 } ),
\end{eqnarray}
where $ {\bf \hat X }=\sum_{i=1}^{n} \sqrt{\hat X_i\hat X_i}$. This algebra induces a maximal  length and a nonzero
minimal momentum. Within this deformed algebra, the action of  position and momentum operators on  the position  space representation can 
be written as
\begin{eqnarray}
	\hat X_i \phi(x)&=&x_i \phi(x) \quad \mbox{and}\quad  \hat P_j \phi(x)=-i\hbar(1-\tau  {\bf  x} +\tau^2  {\bf  x^2})\partial_{x_j}  \phi(x), 
\end{eqnarray}
where $ {\bf  x }=\sum_{i=1}^{n} \sqrt{{ x_i}{ x_i}}$. As in ordinary quantum mechanics,  if we assume that the components of the position operator are commutative
\begin{eqnarray}
	[{\bf\hat X}_i, {\bf\hat X}_j]=0,
\end{eqnarray}
then  from the Jacobi identity
\begin{eqnarray}
	[[{\bf\hat P}_i,{\bf\hat P}_j],{\bf\hat X}_k]+[[{\bf\hat P}_j,{\bf\hat X}_k],{\bf\hat P}_i]+[[{\bf\hat X}_k,{\bf\hat P}_i],{\bf\hat P}_j]=0
\end{eqnarray}
one  obtains the commutation relations between the components of the momentum operator as \cite{11,35,36}
\begin{eqnarray}
	[{\bf\hat P}_i,{\bf\hat P}_j]= i\hbar\tau\left(2\tau-\frac{1}{{\bf  x }}\right) \left({\bf  \hat P}_i{\bf  \hat X}_j-{\bf  \hat P}_i{\bf  \hat X}_j\right).
\end{eqnarray}
The identity operator can be expanded as
\begin{eqnarray}
	\int_{-\infty}^{+\infty}\frac{d^nx}{1-\tau  {\bf  x} +\tau^2  {\bf  x^2}}|{\bf  x}\rangle \langle {\bf  x}|=\mathbb{I}, \label{identity2}.
\end{eqnarray}
The scalar product between two states $|\Psi\rangle$ and $|\Phi\rangle$ and the orthogonality of
eigenstates become
\begin{eqnarray}
	\langle \Psi |\Phi\rangle= \int_{-\infty}^{+\infty}\frac{d^nx}{1-\tau  {\bf  x} +\tau^2  {\bf  x^2}} \Psi^*(x)\Phi(x),\quad 
	\langle {\bf  x} |{\bf  x}'\rangle= (1-\tau  {\bf  x} +\tau^2  {\bf  x^2})\delta^n({\bf  x}-{\bf  x}').
\end{eqnarray}  
Now, if we consider the generalization of the previously defined  as 
\begin{eqnarray}\label{alg1}
	{[{\bf\hat X}_i,{\bf\hat P}_j]}=i\hbar \delta_{ij} \left( \mathbb{I}- f({\bf\hat X})+ g({\bf\hat X^2})\right),
\end{eqnarray}
where $f$ and $g$ are  deformed  functions that  we assume  strictly positive. Then it is straightforward to show that 
\begin{eqnarray}
	{\bf\hat X}_i \phi(x)=x_i \phi(x) \quad \mbox{and}\quad  {\bf\hat P}_j \phi(x)=-i\hbar\left(1-f({\bf  x})+ g({\bf  x^2})\right) \partial_{x_j}  \phi(x).
\end{eqnarray}
Therefore we find \cite{11}
\begin{eqnarray}
[{\bf\hat P}_i,{\bf\hat P}_j]= i\hbar\left(-\frac{1}{{\bf  x }}f'({\bf  x})+ g'({\bf  x^2})\right) \left({\bf  \hat P}_i{\bf  \hat X}_j-{\bf  \hat P}_i{\bf  \hat X}_j\right).
\end{eqnarray}
where by definition
\begin{eqnarray}
f'({\bf  x})=\frac{df}{d{\bf  x}} \quad \mbox{and}\quad g'({\bf  x^2})=\frac{dg}{d{\bf  x^2}}.
\end{eqnarray}
In our case (\ref{algn}), $ f({\bf\hat X})=\tau {\bf\hat X}$ and $ g({\bf\hat X}) =\tau^2 {\bf\hat X^2}$. If we choose $f({\bf\hat X})=\mathbb{I}-\frac{\mathbb{I}}{\mathbb{I}-\alpha {\bf\hat X^2}}$ and $ g({\bf\hat X^2})=0$, we obtain the generalized form of Perivolaropoulos's algbra \cite{1} recently determined by Bensalem and Bouaziz \cite{5}.


\section{Conclusion}
In this paper we have studied a new generalized  form of GUP (\ref{uncertitude}) introduced by  Perivolaropoulos \cite{1}  which implies a maximal length uncertainty and a minimal  observable momentum. Then, using  the deformed algebra (\ref{al1})  which generates this GUP, we studied  the representation of the wave function that   describes both uncertainty measurements. We have shown that with this  maximal length uncertainty  concept
there is no divergency in energy spectrum of
the particle. The   state representations of this particle are   physically meaningfull and  exhibit proprieties similar to the standard Gaussian states which are consequences of quantum fluctuations at that scale. We have  also studied the quasi-representation of operators through the Fourier transform  and its inverse. Finally, we have  generalised the  algebra (\ref{al1}) into n-dimensional cases where  a general representation of operators was  studied.

 \section*{Acknowledgements}
 B. A. Appiah acknowledges the receipt of the grant from AIMS-Ghana. L.M. Lawson would like to thank
  AIMS-Ghana for extremely kind  hospitality during the completion of this work. Thanks for   S. Bensalem for providing us some materials which considerably improved the quality of this paper.

\end{document}